\magnification=\magstep1
\hfuzz=6pt
\baselineskip=15pt

$ $

\vskip 1in

\centerline{\bf The power of entangled quantum channels}

\bigskip

\centerline{Seth Lloyd}

\centerline{MIT Department of Mechanical Engineering}

\bigskip
{\bf 
All communication channels are at bottom quantum mechanical.
Quantum mechanics contributes both obstacles to communication in the 
form of noise, and opportunities in the use of intrinsically
quantum representations for information.  This paper investigates the
trade-off between power and communication rate for coupled quantum channels.
By exploiting quantum correlations such as entanglement, 
coupled quantum channels can communicate at a potentially higher 
rate than unentangled quantum channels given the same power.
In particular, given the same overall power,
$M$ coupled, entangled quantum channels can send $M$ bits
in the same time it takes a single channel to send a single bit,
and in the same time it takes $M$ unentangled channels to send
$\sqrt M$ bits. 
}   

As communication technologies push down to the quantum level, a 
considerable effort has been made to uncover the physical limits to
the communication process.  Much progress has been made in investigating
the physical limits to the bosonic communications channel (1-22) (e.g.,
quantum optical communication (1)) and to quantum communications in
general (23-26).  Quantum systems can be correlated with eachother
in ways that classical systems cannot, a feature known as entanglement.  
It has been speculated that entanglement might be used
enhance the capacity of quantum channels as compared to classical
channels (24-28).  Perhaps the best known example of the use of entanglement
to enhance communication capacity is that of super-dense coding (25),
in which two parties who initially possess an entangled state can send two bits
of classical information by sending a single quantum bit.
In addition, shared prior entanglement may enhance the transmission 
capacity of quantum channels in the presence of noise (28).
In general, the enhancement in channel capacity that can be attained
using entanglement is not known.  This letter investigates the situation 
in which channels are coupled via a nonlinear dynamics to induce an entangled 
state in the process of transmission.  For fixed power,  I show that 
$M$ coupled, entangled quantum channels can transmit information at a rate
$\sqrt M$ times greater than $M$ uncoupled, unentangled channels.  

To see that quantum mechanics can in principle 
attain large gains in channel capacity for power-limited parallel 
channels by entangling them, first review the case of unentangled
parallel quantum channels.   In particular, it is well established
(1-22) that the broadband bosonic channel with power $P$ can transmit 
no more than $C_{1}=\alpha\sqrt{P/\hbar}$ bits per second, where $\alpha =
\sqrt{\pi/3}(1/\ln2)$.  As a 
consequence (1), if the power is spread amongst $M$ unentangled broadband
bosonic channels, each with power $P/M$, the rate of communication
is $C_M^C= \sqrt{M} C_{1}$.  By contrast,
as I now show, $M$ entangled quantum channels using power $P$ can send 
information from A to B at a rate 
$ C_M^Q = \beta M \sqrt{P/\hbar} \approx 
M C_{1} = \sqrt{M}C_M^C, $ 
where $\beta = \sqrt{2/\pi(1-2^{-M})}$.  
That is, for a fixed power,
entangled channels can in principle outperform unentangled channels
by a wide margin for large $M$.  Perhaps more remarkably, $M$ entangled
channels can send $M$ bits in the same time and using the same overall 
power that it takes a single channel to send a single bit.  A summary
of these results is shown in figure 1.  Not surprisingly, 
producing the necessary entangling dynamics for $M$ quantum channels is likely 
to prove difficult.  As will be shown, however, simple demonstrations
of the power of a small number of entangled channels can be performed 
using existing techniques of quantum information processing.  And since 
the potential gains are large, the attempt to entangle $M$ channels may 
be worth the effort.

Quantum channels are physical systems linking sender and receiver.
A quantum channel can be a fiber-optic cable, a wire connecting
two bits in a computer, or a tunneling barrier connecting
two quantum bits in a quantum computer.
To calculate their limits requires a treatment of the channels' dynamics,
together with their interaction with sender and receiver. 
The calculation of the limits to the bosonic channel is detailed and relies
on the physics of quantum electrodynamics (1).  
Here, a simpler and easily generalizable channel model 
is analyzed --- the `qubit' channel.

The qubit channel transmits a quantum bit from A to B.
Suppose that A and B each possess a two-state quantum system, or `qubit.'
A's qubit holds the quantum state $|\psi\rangle$ which is to be
transmitted to B, whose qubit is initially in the state $|0\rangle$.
The two states $|0\rangle$ and $|1\rangle$ of the qubits are assumed
to be degenerate, so that no energy is required to store the qubit.
The qubit channel can be used either to transmit classical information 
-- $|\psi\rangle  = |0\rangle$ or $|1\rangle$ -- 
or to transmit quantum information
-- $|\psi\rangle = \alpha|0\rangle + \beta |1\rangle$.  
The dynamics of the channel should transfer the information from A's
qubit to B's qubit.  After the transfer has taken place, B's qubit
is in the state $|\psi\rangle$ and A's qubit is in a standard state
such as $|0\rangle$.  The sole restriction placed on 
dynamics of the channel is that it obeys the rules of quantum mechanics:
the time evolution of A's qubit and B's qubit, together with their
environment and whatever interaction they use to transfer the information,
is a unitary, Hamiltonian dynamics.  (Note that an arbitrary strictly
positive time evolution can be embedded in such dynamics (29).)  This
requirement alone allows the establishment of a bound for the rate
of information transmission down the qubit channel given limited power.

To derive this bound, note that the average overlap 
$|\langle 0| \psi\rangle|^2$ of A and B's initial
state with their final state is $1/2$.   Accordingly, each time the
channel is used, the total state of A and B together with environment
must evolve by an average angle in Hilbert space of at least $\pi/2$.  That
is, to transmit a 0 requires no transformation of B's qubit, but to
transmit a 1, B's qubit must be rotated by $\pi$.  The 
Margolus-Levitin theorem then implies that when B's qubit is rotated
by $\pi$, the average energy of the complete system above its ground state is
$E\geq \pi \hbar/2 \Delta t$, where $\Delta t$ is the time over which the 
transfer takes place (30).  If a 1 is sent half the time,
the average power associated with applying
energy  $E$ for time $\Delta t$ is $P=E/\Delta t \geq
\pi \hbar/4 \Delta t^2$.  Consequently, the rate at which 
a bit can be reliably transferred from A to B is 
$$ C_{1q} = 1/\Delta t = (2/\sqrt{\pi})
\sqrt{ P/\hbar}.\eqno(1)$$
Equation (1) applies to the reliable transmission of a single
bit.  If one accepts unreliable or noisy transfer, then the
transfer time can be less than $\Delta t = \pi \hbar/4 E$, as
one does not have to rotate the state of A and B by the full angle
$\pi$ to transform them into an orthogonal state.  
This feature, together with the use of error correcting codes,
can be used to enhance the rate of transfer of information for a given
energy (31).   In addition, if one is willing to send less than a full
bit of information by sending a 0 with a higher probability, one
can decrease the energy per transmission time $\Delta t$.

The first obvious point to note about equation (1) is that the 
power-limited capacity of the qubit channel is very similar to
the power-limited capacity of the broadband bosonic channel.  The
capacity of the qubit channel and the bosonic channel differ by 
only a constant of order unity.  At first, this
might seem surprising.  After all, the broadband bosonic channel
has an infinite number of temporal modes, each of which in turn
has an infinite number of states.  In contrast, the qubit channel
has only two qubits, each of which has only two states.  The capacity
of the bosonic channel is attained in the limit of infinite time by
preparing energy eigenstates with a thermal distribution that satisfies
the integrated power constraint.  In contrast,
as will be seen below, the qubit channel operates
by sending one bit dynamically in time $\Delta t$ while respecting
the power constraint.  The two channels -- bosonic and qubit --
could hardly look more different.  Why then do they exhibit almost
the same channel capacity, including the same dependence on power?

The answer is that both of these channel capacities 
are attained by creating distributions
of states with the maximum spread in energy given the overall power 
constraint --- they are both broadband systems (1, 30).  
In addition, the bosonic channel attains its maximum capacity
in the limit where approximately one bit is transmitted for
each quantum sent down the channel: at its capacity limit,
the bosonic channel closely resembles the qubit channel.  
To see broadband nature of the qubit channel, 
examine how the channel capacity limit can be attained.  

A simple example of an interaction that attains
the qubit capacity limit is the application of a `swap'
operation: $S= \sum_{ij=0}^1 |ij\rangle_{AB}\langle ji|$.
$S$ is a unitary transformation that swaps the quantum information
in A's qubit with the quantum information in B's qubit.
Note that $S^2 = 1$: two applications of $S$ returns the bits
to their original states.  Consequently,$S$ is 
Hermitian and has eigenvalues $\pm 1$.  Note also that
$e^{-i\theta S} =  \cos\theta -i \sin \theta S$.

Now analyze the power needed to apply a swap operation.
Apply the Hamiltonian $\tilde S = \pi\hbar(1-S)/2\Delta t$
to A and B's qubits 
for time $\Delta t$.  It is straightforward to verify that
$e^{-i\tilde S\Delta t/\hbar} = S$.
So applying $\tilde S$ for time $\Delta t$ swaps the qubits.
The average energy of A and B and the apparatus that swaps
the qubits is 
$$E= (\langle \psi |_A \langle 0|_B)
\tilde S (|0\rangle_A |\psi\rangle_B) =  
\pi\hbar(1-|\langle 0|\psi\rangle|^2)/2\Delta t. \eqno(2)$$
Averaging over states $|\psi\rangle$ gives an energy $E$ that 
saturates the Margolus-Levitin bound.  Swap attains the
power/capacity limit of equation (1) above. 

The `swap' picture of quantum information transmission assumes a
direct transfer of A's qubit to B.  A similar
picture holds in which A's and B's qubits are coupled by an intervening
chain of qubits $A_1B_1 A_2 B_2 \ldots A_nB_n$, where A has access
to $A_1$ and B has access to $B_n$.  Here, quantum information can
be sent along the chain by swapping $A_i$ with $B_i$ over a time
$\Delta t/2$, then swapping $B_i$ with $A_{i+1}$, and repeating until
the qubit has been moved from A to B.  In this case, the time taken
to send a qubit from A to B is $n\Delta t$, and the average energy
employed is $2nE$, giving an average power of $2P=2E\Delta t$.  
The rate at which information is sent down the channel is still 
1 bit in time $\Delta t$.  
Accordingly, the transmission of information from A to B
by repeated swapping down a chain of qubits comes within $\sqrt 2$
of the the power/capacity limit of (1).  In fact, a variety of Hamiltonians
(e.g., $ H= S_{A_1B_1} + S_{B_1A_2} + \ldots S_{A_nB_n}$)
can be used to propagate `spin waves' down the qubit chain
at rates on the order of the power/capacity limit (1).
The similarity of the power/capacity tradeoff for chains
of qubits and for bosons should not be surprising as the physics of 
spin wave propagation is closely related to the physics of wave
propagation in the multi-mode bosonic channel.

Now investigate the case of multiple quantum channels.  It is here
that entanglement leads to a significant enhancement in power-limited
transmission rate.  Clearly, $M$ parallel quantum
channels can transmit information at a rate $\sqrt{M}$ greater than a single
quantum channel using the same power $P$ merely by dividing the power equally
amongst the channels (1).  Each channel now transmits at a rate $(2/\sqrt\pi)
\sqrt{P/M\hbar}$ 
giving an overall rate of transmission $C_{M}^C = (2M/\sqrt\pi) \sqrt{P/Mh} = 
(2/\sqrt\pi) \sqrt{MP/h} = \sqrt M C_1^C$.  This rate enhancement for parallel
unentangled channels holds for both the bosonic channel and
for the qubit channel.  Because of the square root dependence of
transmission rate on power, both the qubit
and broadband bosonic channel are more efficient at a lower power.
As a result, one improves performance by dividing 
up information and power among the different channels. 

If one is able to apply operations that entangle the
channels, one can do even better, as I now show.
The goal of the $M$-channel transfer is to 
enact the $2M$-qubit analog of the swap above:
$$S_{1\ldots M} =  \sum_{i_1j_1\ldots i_Mj_M = 0}^1
|i_1j_1\rangle_{AB_1}\langle j_1i_1| \otimes \ldots \otimes
|i_Mj_M\rangle_{AB_M}\langle j_Mi_M|. \eqno(3)$$
The $2M$ qubit swap  $S_{1\ldots M}$ swaps A's $M$  qubits with 
B's M qubits and has the same properties  as the 2-qubit swap above
(Hermitian, squares to one, etc.).
As above, define the Hamiltonian $\tilde S_{1\ldots M}
= \pi\hbar (1-S_{1\ldots M})/2\Delta t$.
Applying the Hamiltonian $\tilde S_{1\ldots M}$ 
for a time $\Delta t$ then swaps
A's qubit string with B's qubits.  
The average energy during the $M$-qubit swap is
$E= \pi\hbar(1-|\langle 0|\psi\rangle|^2)/2\Delta t,$
as in equation (2) above.  Now, however, 
$|\langle 0|\psi\rangle|^2 = 1/2^M$ for a randomly selected
$|\psi\rangle$.  Accordingly, the time taken to transfer
A's bit to B using power $P$ is given by 
$$1/\Delta t = \sqrt{2(1-2^{-M})P/\pi\hbar},\eqno(4)$$
which is the $M$-channel analog of the limit (1).  
The time taken to perform the transfer using
power $P$ attains the limit (4), but now for the transfer of $M$ bits
rather than a single bit.  

Application of the Hamiltonian $S_{1\ldots M}$
transfers $M$ bits down $M$ parallel qubit
channels using essentially the same energy $E \approx\hbar/\Delta t$, the
same power $P\approx\hbar/\Delta t^2$, and in the same time $\Delta t$
it takes a single channel
to transmit a single bit.  Similar results hold for the transmission
of $M$ qubits down $M$ chains of $n$ qubits, as above: the transmission
time in this case is $n$ times as long, but the power is the same
as the single bit case, while the number
of bits per second is $M$ times the single qubit channel rate.

It is easy to verify that during the transfer, the
$M$ qubit channels are mutually entangled.  For example, if A's input state
is $|b_M\rangle= | b_1\ldots b_M\rangle$, then at time 
$\Delta t/2$ (halfway through the controlled
flipping operation) A and B's qubits are in the state 
$$(e^{-i\pi/4}/\sqrt2)\big( | b_1\ldots b_M\rangle_A
|00\ldots 0\rangle_B + i |00\ldots 0\rangle_A
|b_1\ldots b_M\rangle_B \big).\eqno(5)$$  
The fact that the channels are entangled during the course of
transmission has implications for the sensitivity of the entangled
channel to noise.  On the one hand, entangled states of the form
(5) are typically $\sqrt M$ times more sensitive to decoherence than
unentangled states.  On the other hand, this sensitivity to decoherence
can actually be an advantage in sending information.  Suppose
that $A$ never sends the state $00\ldots 0$.  If the multiple channel
is decohered in the course of transmission, then B either receives
$00\ldots 0$ or the correct message $b_1\ldots b_M$.  If he receives
$00\ldots 0$, he just waits and measures again until he receives
$b_1\ldots b_M \neq 00\ldots 0$.  That is, decoherence on its own
gives no errors for the transmission of classical information down
the entangled channels.  In fact, as noted in (31), decoherence can
actually enhance the rate of transmission for a fixed power.  

Transferring $M$ bits down $M$ unentangled quantum channels 
using the same power as a single qubit
channel takes $\sqrt M$ times longer than transferring the
information down coupled, entangled channels.
Unentangled transfer corresponds to the
application of $M$ two-qubit swap operations with Hamiltonian 
$\tilde S_1 + \ldots \tilde S_M$ as opposed to the $2M$-qubit 
swap Hamiltonian $\tilde S_{1\ldots M}$, and takes $\sqrt M$ times
the energy of the entangled swap.  As a result,   
the coupled, entangled channels have a 
capacity of at least $\sqrt M$ times the capacity of the uncoupled, 
unentangled channels.  

Perhaps the most remarkable aspect of this result is that the use of 
entanglement allows the transfer of $M$ bits in the same time
and using the same power that it takes to transfer
a single bit.  Does entanglement truly allow one to get `something for nothing'
as this result suggests?  What is the catch?  After all, previous
investigations of the use of entanglement in uncoupled quantum
channels have found at best modest increases in channel capacity (23-27)
(with the possible exception of noisy quantum channels (28)).

In fact, entanglement does indeed allow the capacity increase derived above.   
To find the absolute upper bound on the capacity of coupled quantum
channels will require the detailed application of Kholevo's theorem (23, 1).
However, the Margolus-Levitin theorem (30) is very general, and suggests
that the absolute limit on coupled channel capacity differs from (3)
by at most a constant of order unity.  
In a certain sense, that it is just as easy in terms
of power and energy to rotate $2M$ bits from one state to another as it is to
rotate 2 bits from one state to another should not be surprising: no two states
in Hilbert space are more than angle of $\pi$ apart.  Accordingly, if one
can effect arbitrary evolutions on the $M$-qubit channel Hilbert space,
$M$ bits can be transferred using the same power and time as one bit.
The situation is summarized in figure 2.
Effectively, the coupling between the channels allows them to transmit
information in the form a `super-boson' with $2^M$ internal states.  
The $\sqrt M$ enhancement afforded by exploiting entanglement is
typical of quantum information processing and arises from essentially
the same source as the $\sqrt M$ enhancements in quantum search (32) 
and quantum positioning (33).

The catch is that enacting the necessary Hamiltonian $\tilde S_{1\ldots M}$ 
is likely to prove experimentally difficult.  Even for two qubit 
channels, enacting the Hamiltonian of equation (2) involves entangling four
quantum bits, a difficult action using current technologies.  One might 
hope to be able to build up this Hamiltonian time evolution using 
elementary quantum logic operations on two quantum bits at a time, 
but in this case most of the power advantage is lost, as the net angle 
rotated in Hilbert space becomes larger than $\pi$.  To attain the 
$\sqrt M$ enhancement of channel capacity allowed by entanglement, 
an $M$-qubit entangling operation must be used.  Such operations 
correspond to interaction operators of the
form $\sigma_x^1 \sigma_x^2 \ldots \sigma_x^M$ for spin qubits and
$a_{A1} a^\dagger_{B1} \ldots a_{AM} a^\dagger_{BM} + H.C.$ for particle
modes.  That is, $M$'th order nonlinear interactions are required to
attain entanglement-enhanced channel capacity.  Such interactions are
hard to enact experimentally, although it is possible to use simple
quantum logic and quantum communication devices to perform
proof-of-principle demonstrations of entanglement-enhanced capacity for
small $M$ (34-36).  For example, suppose that A wants to use microwaves
or light to load two bits onto the nuclear spins or hyperfine levels
of B's two atoms.  The results derived above show that if A is able
to manipulate entangling interactions between the two spins or atoms, the two
bits can be loaded using $\sqrt 2$ less power than in the case that
the spins or atoms remain unentangled.   Such a proof-of-principle
experiment could be performed using nuclear resonance on two spins
in a molecule, or using optical resonance on two interacting atoms
in a trap.  For larger $M$,
enacting the proper entangling coupling is likely to prove experimentally
difficult, but if such coupling can be enacted,
substantial gains in quantum channel capacity can be obtained.
Whether or not the potential gains afforded by entanglement can
be realized in experimentally feasible quantum optical systems
acting over significant distances remains an open question.

This paper has derived power/speed limits for the problem of 
transferring $M$ qubits reliably from A to B. By use of entanglement,
the $M$ qubits can be transferred $\sqrt M$ times more rapidly for
the same power as $M$ unentangled qubits.  The absolute capacity of
$M$ coupled, entangled qubit channels will have to be derived by
the sophisticated application of Holevo's theorem (27).
Open questions include the fully relativistic treatment
of entangling channels, the possibility of significant but lesser
communication enhancements by the use of partial entanglement,
and the effect of noise on entangling channels.
But as this paper shows, entanglement in principle gives a significant
increase in channel capacity.
 
\vfill
\noindent Acknowledgements: The author would like to thank 
Vittorio Giovannetti, Lorenzo Maccone, Hermann Haus, Jeff Shapiro,
and Franco Wong for inspiration and for helpful discussions.
\vfil\eject

\noindent References:

\bigskip\noindent (1)  C.M. Caves, P.D. Drummond, {\it Rev. Mod.
Phys.} {\bf 66}, 481-537 (1994).

\bigskip\noindent (2) J.P. Gordon, in {\it Advances in Quantum 
Electronics}, J.R. Singer ed., (Columbia, New York, 1961),
p. 509.

\bigskip\noindent (3) D.S. Lebedev, L.B. Levitin, {\it Dok. Akad.
Nauk SSSR} {\bf 149}, 1299; {\it Sov. Phys. Dokl.} {\bf 8}, 377
(1963).

\bigskip\noindent (4) D.S. Lebedev, L.B. Levitin, {\it Inf. Control}
{\bf 9}, 1 (1966).

\bigskip\noindent (5) H. Marko, {\it Kybernetik} {\bf 2}, 274 (1965).

\bigskip\noindent (6) H. Takahasi, in {\it Advances in Communication
Systems, Vol. 1}, A.V. Balakrishnan, ed., (Academic, New York 1965), 
p. 227.

\bigskip\noindent (7) J.I. Bowen, {\it IEEE Trans. Inf. Theory}
{\bf IT-13}, 230 (1967).

\bigskip\noindent (8) R. Landauer, J.W.F. Woo, in {\it Synergetics},
H. Haken, ed., (Tuebner, Stuttgart 1973), p. 97.

\bigskip\noindent (9) C.W. Helstrom, {\it Proc. IEEE} (Lett.) {\bf 62},
139 (1974).

\bigskip\noindent (10) H.P. Yuen, J.H. Shapiro, {\it IEEE Trans. Inf. 
Theory} {\bf IT-24}, 657 (1978).

\bigskip\noindent (11) J.H. Shapiro, H.P. Yuen, J.A. Machado Mata,
{\it IEEE Trans. Inf. Theory} {\bf IT-25}, 179 (1979).

\bigskip\noindent (12) H.P. Yuen, J.H. Shapiro, {\it IEEE Trans. Inf. 
Theory} {\bf IT-26}, 78 (1980).

\bigskip\noindent (13) L.B. Levitin, {\it Int. J. Th. Phys.} {\bf 21},
299 (1982).

\bigskip\noindent (14) J.B. Pendry, {\it J. Phys. A} {\bf 16}, 2161 (1983).

\bigskip\noindent (15) Y. Yamamoto, H.A. Haus, {\it Rev. Mod. Phys.} {\bf 58},
1001 (1986).

\bigskip\noindent (16) B.E.A. Saleh, M.C. Teich {\it Phys. Rev. Lett.}
{\bf 58} 2656 (1987).

\bigskip\noindent (17) R. Landauer, {\it Nature 335}, 779 (1988).

\bigskip\noindent (18) J.D. Bekenstein, {\it Phys. Rev. A} {\bf 37}, 3437
(1988).

\bigskip\noindent (19) J.D. Bekenstein, M. Schiffer, {\it Int. J. Mod.
Phys. C} {\bf 1}, 355 (1990).

\bigskip\noindent (20) M. Schiffer, {\it Phys. Rev. A} {\bf 43}, 5337 (1991).

\bigskip\noindent (21) H.P. Yueh, M. Ozawa, {\it Phys. Rev. Lett.} {\bf 70}.
363 (1992).

\bigskip\noindent (22) B.E.A. Saleh, M.C. Teich, {\it Proc. IEEE} {\bf 80},
451 (1992). 

\bigskip\noindent (23) A.S. Kholevo, {\it Probl. Peredachi Inf.} {\bf 9},
3 (1973) [{\it Probl. Inf. Transm. (USSR)} {\bf 9}, 177 (1973)].

\bigskip\noindent (24) E.H. Lieb, M.B. Ruskai, {\it J. Math. Phys.}
{\bf 14}, 1938 (1973).

\bigskip\noindent (25) C.H. Bennett, S.J. Wiesner, {\it Phys. Rev. Lett.}
{\bf 69}, 2881 (1992).

\bigskip\noindent (26) B. Schumacher, M.D. Westmoreland, {\it Phys. Rev.
A} {\bf 56}, 131 (1997).

\bigskip\noindent (27) A.S. Holevo, {\it IEEE Trans. Inf. Theory}
{\bf 44}, 269 (1998). 

\bigskip\noindent (28) C.H. Bennett, P.W. Shor, J.A. Smolin, A.V. Thapliy,
`Entanglement-Assisted Capacity of a Quantum Channel
and the Reverse Shannon Theorem,' arXiv: quant-ph/0106052.

\bigskip\noindent (29) A. Peres, {\it Quantum Theory:
Concepts and Methods}, (Kluwer, Hingham) 1995.
 
\bigskip\noindent (30) N. Margolus, L.B. Levitin, L.B., in 
{\it PhysComp96}, T. Toffoli, M. Biafore, J. Leao, eds. 
(NECSI, Boston) 1996; {\it Physica D} {\bf 120}, 188-195 (1998).

\bigskip\noindent (31) A. Harrow, S. Lloyd, to be published.

\bigskip\noindent (32)
Grover, L.K., in {\it Proceedings of the 28th Annual ACM Symposium
on the Theory of Computing}, ACM Press, New York, 1996, pp. 212-218.

\bigskip\noindent (33)
V. Giovannetti, S. Lloyd, L. Maccone, {\it Nature} (2001).

\bigskip\noindent (34)
Turchette, Q.A., Hood, C.J., Lange, W., Mabuchi, H.,
Kimble, H.J., {\it Phys. Rev. Lett.}, {\bf 75}, 4710-4713 (1995).

\bigskip\noindent (35) Monroe, C., Meekhof, D.M., King, B.E., Itano, W.M.,
Wineland, D.J., {\it Phys. Rev. Lett.}, {\bf 75}, 4714-4717 (1995).

\bigskip\noindent (36) Cory, D.G., Fahmy, A.F., Havel, T.F.,
in {\it PhysComp96, Proceedings of the
Fourth Workshop on Physics and Computation}, T. Toffoli, M. Biafore,
J. Le\~ao, eds., (New England Complex Systems Institute, Boston) 1996.

\bigskip\noindent (37) Gershenfeld, N.A., Chuang, I.L.
{\it Science} {\bf 275}, pp. 350-356
(1997).

\vfill\eject
\noindent Figure captions:

\bigskip\noindent Figure 1: Summary of the power/capacity tradeoff
for entangled and unentangled channels.  

1a) A single channel can transmit information at a rate 
$\approx\sqrt{ P/\hbar}$, where $P$ is the power.

1b) Because the channels are more efficient at low power, $M$ unentangled
channels can send information at a rate $\approx M\sqrt{P/M\hbar} = \sqrt M
\sqrt{P/\hbar}$.

1c) If the channels are coupled to entangle them in the course of
transmission, they can send information at a rate 
$\approx M\sqrt{P/\hbar}.$

\bigskip\noindent Figure 2: Explanation of the $\sqrt M$
enhancement from entanglement.  No two states in Hilbert space
are more than an angle $\pi$ apart.  As a result,
to send a state $|\psi\rangle$ from A to B in time $\Delta t$
requires energy $E\approx\hbar/\Delta t$, independent
of the number of bits $M$ in $|\psi\rangle$.  The direct path
through Hilbert space passes through states in which
the $M$ channels are maximally entangled.
In contrast, the transmission path corresponding
to unentangled states is $\sqrt M$ times longer.

\vfill\eject
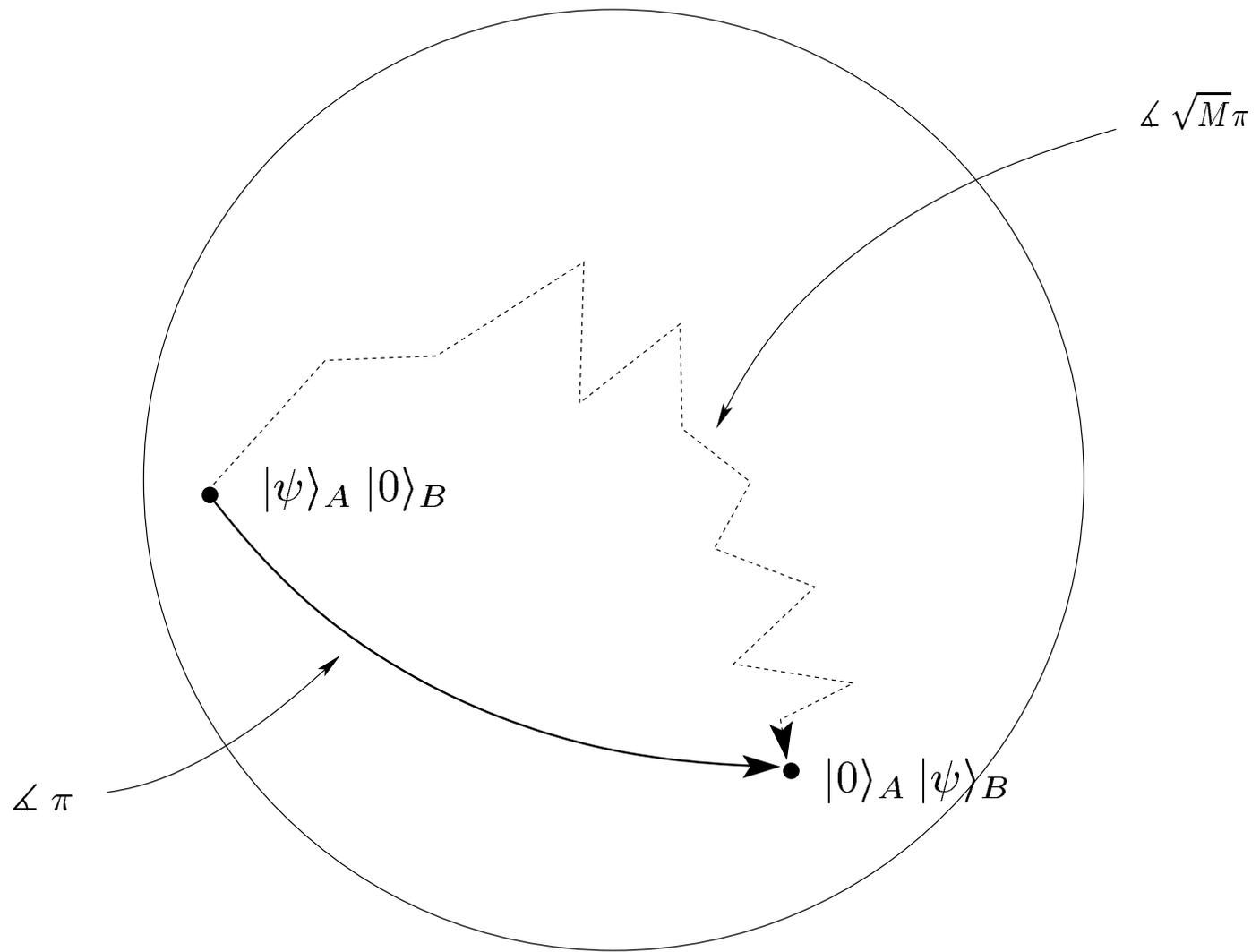
\end